\documentclass[aps,showpacs,11pt]{revtex4}
\usepackage{dcolumn}
\usepackage{graphicx}
\usepackage{amsmath}
\usepackage{amsfonts}
\usepackage{amssymb}
\usepackage{psfrag}
\usepackage{wrapfig}
\usepackage{subfigure}
\usepackage{makeidx}
\usepackage{bm}
\usepackage{epsf}
\usepackage{hyperref}
\makeatletter

\newcommand{\Rmnum}[1]{\expandafter\@slowromancap\romannumeral #1@}
\makeatother

\begin{document}

\title{Thermodynamics of rotating black holes with scalar hair in three dimensions}

\author{De-Cheng Zou}
\email{zoudecheng@sjtu.edu.cn}
\author{Yunqi Liu}
\email{liuyunqi@sjtu.edu.cn}
\author{Bin Wang}
\email{wang$_$b@sjtu.edu.cn}

\affiliation{Department of Physics and Astronomy, Shanghai Jiao Tong University, Shanghai 200240, China}

\author{Wei Xu}
\email{xuweifuture@gmail.com}

\affiliation{School of Physics, Huazhong University of Science and Technology, Wuhan 430074, China}

\date{\today}

\begin{abstract}

Introducing a new form of scalar potential $V(\phi)$,
we derive a proper form of the rotating black hole solution
in three-dimensional Einstein gravity with nonminimally coupled scalar field
and find that the first law of thermodynamics of this new rotating
hairy black hole can be protected,
where the scalar field parameter $B$ is constrained to relate
to the black hole size. We also disclose the
Hawking-Page phase transition between this rotating hairy
black holes and the pure thermal radiation. Moreover,
we study phase transitions between this rotating hairy black
hole and rotating BTZ black hole. Considering the matchings for
the temperature and angular momentum, we find that
the rotating BTZ black hole always has
smaller free energy which is a thermodynamically
more preferred phase. Additionally, we evaluate the thermodynamics of the
rotating black hole with minimally coupled scalar hair in
three dimensions, which exhibits that the thermodynamical
behaviors of this rotating hairy black hole
are very similar to those of the rotating black hole
with nonminimally coupled scalar hair.
\end{abstract}


%\keywords{black hole, scalar field, three dimensions, phase transition}

\maketitle

\section{Introduction}

Hairy black holes are interesting solutions in
gravitational theories. They have been studied
extensively over the past few years, mainly in
connection with the no hair theorem. In general,
the no-hair theorem rules out four-dimensional
black holes coupled to scalar field in
asymptotically flat spacetimes
\cite{Bocharova,Bekenstein:1974sf,Bekenstein:1975ts}
because they are not physically acceptable since
the scalar field can diverge on the horizon and
black holes can become unstable
\cite{Bronnikov:1978mx}. In higher-dimensional
cases ($D>4$), the hairy black hole solutions in
asymptotically flat spacetimes  simply do not
exist \cite{Xanthopoulos:1992fm,
Klimcik:1993cia}. When the cosmological constant
is taken into account, the no-hair theorem can
usually be circumvented, and then some hairy
black holes
can be found and the divergence of
the scalar field can be hidden behind the
horizon. In asymptotically de Sitter (dS)
spacetimes, hairy black hole minimally coupled
to a scalar field were presented in Ref.~
\cite{Zloshchastiev:2004ny}, but it was found
dynamically unstable. The exact black hole
solutions with nonminimally coupled scalar
field in dS spacetime have also been discussed
in Refs.~\cite{Martinez:2002ru,Barlow:2005yd}, while
they were also found unstable
\cite{Harper:2003wt,Dotti:2007cp}.

When a negative cosmological constant is
considered, four-dimensional (MTZ) black hole
solutions with minimal
\cite{Martinez:2004nb,Martinez:2006an} and
nonminimal \cite{Martinez:2005di} scalar fields
can be obtained. Thermodynamically, the phase
transitions between the MTZ black holes and black
holes without scalar hair have been found of the
second order both in the canonical ensemble
\cite{Myung:2008ze} and the grand canonical
ensemble \cite{Martinez:2010ti}. The hairy black
holes in anti-de Sitter(AdS) spaces were found
stable by examining quasinormal modes in their
backgrounds
\cite{Koutsoumbas:2006xj,Shen:2007xk}.
Interestingly, the behavior of quasinormal modes
was disclosed useful to reflect the
thermodynamical second-order phase transition
\cite{Koutsoumbas:2006xj,Shen:2007xk}.
More discussions on the hairy black hole
solutions in AdS spacetimes can be
found in Refs.~
\cite{Winstanley:2005fu,Kolyvaris:2009pc,Gonzalez:2013aca,Anabalon:2012tu,Nadalini:2007qi,
Feng:2013tza,Lu:2014maa,Dias:2011tj}. Further
explorations of the AdS hairy black holes in
higher-curvature gravity theories have been
reported in
\cite{Gaete:2013ixa,Correa:2013bza,Gaete:2013oda,Giribet:2014bva}.
In three dimensional Einstein gravity, some static
black holes with minimally \cite{Henneaux:2002wm} or nonminimally
\cite{Martinez:1996gn,Nadalini:2007qi} coupled
scalar hairs were found. Moreover, the thermodynamical
properties of these hairy black holes were
studied in
\cite{Myung:2008ze,Banados:2005hm,Myung:2006sq},
where the first law of black hole thermodynamics
was found always hold. It is interesting to note that the rotating
black holes with nonminimally
\cite{Zhao:2013isa} and minimally
\cite{Xu:2014uha} coupled scalar hairs in
the three dimensional Einstein gravity have been recently also obtained.
Examining the rotating
black hole with non-minimally coupled scalar field in three
dimensions \cite{Zhao:2013isa},
however, one can find that the thermodynamical properties of this rotating
black hole were incorrectly presented in \cite{Belhaj:2013ioa},
where the first law of black hole thermodynamics cannot be protected when
the parameter $B$ of the scalar field is completely free.
This is the first motivation of the present paper.
On the other hand, the thermodynamical quantities
$M$ and $J$ of this rotating hairy black hole also emerge in
the expression for the scalar potential $V(\phi)$ \cite{Zhao:2013isa},
which makes the action to be non-invariant.
How to overcome these obstacles ? It is interesting to pursue.

In this article, we will first set a new form of the scalar
potential $V(\phi)$, which refers to three parameters $\mu$, $\alpha$ and $l$,
and obtain a proper form of the rotating black
hole solution in the three-dimensional Einstein gravity
with nonminimally coupled scalar field.
After the discussion for the black hole horizon structures, the
parameter $B$ of the scalar field $\phi(r)$ will
be described to be related to the black hole horizon size through
$r_+=\theta\times B$, where $\theta$ is a
dimensionless constant. Fortunately, it will show that this
constraint on the parameter $B$ is required to
ensure the validity of the first law of
thermodynamics for this rotating hairy black hole with
nonminimally coupled scalar field.
Furthermore, the phase transition between the
three-dimensional rotating black hole
with a nonminimally coupled scalar field and the
rotating BTZ black hole will be discussed
by using the temperature and angular momentum matchings.
Besides, we will also extend these investigations to
the rotating black hole with minimally coupled to scalar
field in three dimensions \cite{Xu:2014uha}, and
discuss the phase transition between these
rotating hairy and BTZ black holes.

This paper is organized as follows. In
Sec.~\ref{2s}, we will present a proper form of a rotating black
hole solution in three-dimensional Einstein gravity with
nonminimally coupled scalar field. In
Sec.~\ref{3s}, we will discuss the
thermodynamical properties of this rotating hairy
black hole and examine the first law of black
hole thermodynamics. Moreover, by adopting temperature
and angular momentum matchings, we will study the
phase transition between the rotating hairy black
hole and rotating BTZ black hole. In
Sec.~\ref{4s}, we will extend the discussion to
the  rotating black hole with minimally coupled
scalar hair. Finally, we will present
conclusions and discussions in Sec.~\ref{5s}.

\section{Rotating black holes nonminimally coupled to scalar field}
\label{2s}

We start with the action in three-dimensional
Einstein gravity with a nonminimally coupled
scalar field \cite{Zhao:2013isa}
\begin{eqnarray}
{\cal I}=\frac{1}{2}\int{d^3x\sqrt{-g}\left(R-g^{\mu\nu}\nabla_{\mu}\phi\nabla_{\nu}\phi
-\xi R\phi^2-2V(\phi)\right)},\label{i}
\end{eqnarray}
where $\xi$ equals to $1/8$ signifying the coupling strength between gravity and the scalar field.
Taking the scalar potential \cite{Zhao:2013isa}
\begin{eqnarray}
V(\phi)=-\frac{1}{l^2}+\frac{1}{512}\left(\frac{1}{l^2}+\frac{\beta}{B^2}\right)\phi^6
+\frac{1}{512}(\frac{a^2}{B^4})\frac{\left(\phi^6-40\phi^4
+640\phi^2-4608\right)\phi^{10}}{(\phi^2-8)^5},\label{Vr}
\end{eqnarray}
the rotating black hole solution is given by
\begin{eqnarray}
&&ds^2=-f(r)dt^2+f(r)^{-1}dr^2+r^2\left(d\psi+\omega(r)dt\right)^2,\label{met}\\
&&f(r)=3\beta+\frac{2B\beta}{r}+\frac{\left(3r+2B\right)^2a^2}{r^4}+\frac{r^2}{l^2},
\quad \omega(r)=-\frac{\left(3r+2B\right)a}{r^3}, \label{fr}
\end{eqnarray}
and the scalar field takes the form
\begin{eqnarray}
\phi(r)=\pm\sqrt{\frac{8B}{r+B}}.\label{phi}
\end{eqnarray}
Here, the parameters $\beta$, $B$, and $a$ are
integration constants, and $\Lambda=\frac{1}{l^2}$ appears in $V(\phi)$
as a constant term, which plays the role of a (bare) cosmological constant.

The hairy black hole solution [Eq.~(\ref{fr})]
can reduce to the rotating BTZ black hole
solution \cite{Banados:1992wn} if one takes $B\rightarrow0$, which
leads to $\beta=-\frac{M}{3}$ and $a=\frac{J}{6}$.
The rotating black hole solution [Eq.(4)] can be
rewritten into \cite{Zhao:2013isa}
\begin{eqnarray}
f(r)=-M\left(1+\frac{2B}{3r}\right)+\frac{r^2}{l^2}+\frac{\left(3r+2B\right)^2J^2}{36r^4},\quad
\omega(r)=-\frac{\left(3r+2B\right)J}{6r^3}.\label{fa}
\end{eqnarray}
Nevertheless, it is necessary to point out that the action [Eq.(1)] is not invariant
due to the existence of the thermodynamical quantities
$M$ and $J$ of this rotating hairy black hole in
the scalar potential $V(\phi)$.

The thermodynamical quantities of this rotating hairy
black hole are \cite{Belhaj:2013ioa}
\begin{eqnarray}
M&=&\frac{J^2l^2\left(2B+3r_+\right)^2+36r_+^6}{12l^2r_+^3\left(2B+3r_+\right)},\label{ma}\\
T&=&\frac{f'(r_+)}{4\pi}=\frac{(B+r_+)\left[36r_+^6-J^2l^2(2B
+3r_+)^2\right]}{24\pi l^2 r_+^5\left(2B+3r_+\right)},\label{ta}\\
S&=&\frac{A_H}{4G}\left(1-\xi \phi^2(r_+)\right)=\frac{4\pi r_+^2}{B+r_+},\label{sa}\\
\Omega_H&=&-\omega(r_+)=\frac{\left(3r_{+}+2B\right)J}{6r_+^3},
\end{eqnarray}
where the parameter $B$ of the scalar field is
set to be arbitrary. One can also compute
\begin{eqnarray}
\frac{\partial M}{\partial S}=\frac{(B+r_+)^3\left[36r_+^6-J^2l^2(2B
+3r_+)^2\right]}{8\pi l^2 r_+^5\left(2B+3r_+\right)^2\left(2B+r_+\right)},
\end{eqnarray}
which is different from Eq.~(\ref{ta}), so
that the first law of black hole thermodynamics
meets the challenge in  this rotating hairy black
hole.

Now, we consider a new form of the scalar potential $V(\phi)$
\begin{eqnarray}
V(\phi)=-\frac{1}{l^2}+\frac{1}{512}\left(\frac{1}{l^2}+\mu\right)\phi^6
+\frac{\alpha^2\left(\phi^6-40\phi^4+640\phi^2-4608\right)\phi^{10}}{512(\phi^2-8)^5},\label{Vn}
\end{eqnarray}
where $\mu$, $\alpha$, and $l$ are constant parameters.
Performing the dimensional analysis, we have $[l]=L$,
$[\mu]=L^{-2}$ and $[\alpha]=L^{-1}$.
Moreover, the qualitative behavior of the scalar potential (Eq.(2))
has been adequately analyzed in \cite{Zhao:2013isa}.
Inserting Eq.~(\ref{Vn}) into the action, we can
find the rotating hairy black hole with the
metric coefficient
\begin{eqnarray}
f(r)=\mu B^2\left(3+\frac{2B}{r}\right)+\frac{\left(3r
+2B\right)^2\alpha^2B^4}{r^4}+\frac{r^2}{l^2},\quad
\omega(r)=-\frac{\alpha B^2\left(3r+2B\right)}{r^3}.\label{nf}
\end{eqnarray}
Now, when $\alpha\rightarrow 0$,
the scalar potential $V(\phi)$ and
rotating black hole solution $f(r)$ reduce to the static counterparts
of the static black hole solution in three-dimensional
Einstein gravity with a nonminimally coupled
scalar field \cite{Nadalini:2007qi}.

To see the horizon structures of this proper form of
rotating hairy black hole,  we define
\begin{eqnarray}
X=\frac{3r_{+}+2B}{r_+^3}, \label{x}
\end{eqnarray}
so that $f(r_+)=0$ gives
\begin{eqnarray}
\alpha^2 B^4 X^2+\mu  B^2 X+\frac{1}{l^2}=0,
\end{eqnarray}
which leads to
\begin{eqnarray}
&&X_1=\frac{-\mu-\sqrt{\mu^2-4\alpha^2/l^2}}{2\alpha^2B^2}=\frac{\tilde{X}_1}{B^2}, \nonumber\\
&&X_2=\frac{-\mu+\sqrt{\mu^2-4\alpha^2/l^2}}{2\alpha^2B^2}=\frac{\tilde{X}_2}{B^2}.
\end{eqnarray}
Here $X_{\nu}, (\nu=1,2)$ are both real positive
when $-\mu\geq\frac{2\alpha}{l}$, but they are
both imaginary when $-\mu<\frac{2\alpha}{l}$. We
focus on the case in which
$-\mu\geq\frac{2\alpha}{l}$; thus,
$\tilde{X}_{\nu}>0$.

Equating $X$ to  $X_\nu$, Eq.~(\ref{x}) becomes
\begin{eqnarray}
H_{\nu}(r_+)=X_{\nu}r_+^3-3r_{+}-2B=\tilde{X}_{\nu}r_+^3-3B^2r_{+}-2B^3=0. \label{xx}
\end{eqnarray}
The only positive solution of Eq.~(\ref{xx})
exists when
$\frac{108B^6\left(\tilde{X}_{\nu}-1\right)}{\tilde{X}_{\nu}^3}>0$,
namely $\tilde{X}_{\nu}\geq1$, which is given by
\begin{align}
  r_+^{\nu}=\frac{1}{\tilde{X}_{\nu}}\left[\left(\tilde{X}_{\nu}^2
  -\sqrt{\left(\tilde{X}_{\nu}-1\right)\tilde{X}_{\nu}^3}\right)^{1/3}
  +\left(\tilde{X}_{\nu}^2-\sqrt{\left(\tilde{X}_{\nu}-1\right)\tilde{X}_{\nu}^3}\right)^{1/3}\right]B
  =\hat{\theta}(\tilde{X}_{\nu}) B.\label{root}
\end{align}
When
$\frac{108B^6\left(\tilde{X}_{\nu}-1\right)}{\tilde{X}_{\nu}^3}<0$,
namely, $0<\tilde{X}_{\nu}<1$, there exist three
roots for the solution of Eq.~(\ref{xx}), but we
only have interest in the positive one
\begin{align}
 r_+^{\nu}=\frac{B}{\sqrt{\tilde{X}_{\nu}}}\left(\cos\eta+\sqrt{3}\sin\eta\right)
 =\bar{\theta}(\tilde{X}_{\nu}) B,\quad
 \eta=\frac{\arccos(-\sqrt{\tilde{X}_{\nu}})}{3}.
\end{align}
From the horizon structure, we find the
constraint condition for the parameter of the
scalar field $B$,
$r_+^{\nu}=\theta(\tilde{X}_{\nu}) B$ when
$\tilde{X}_{\nu}>0$. Then $\theta(\tilde{X}_{\nu})$
is a dimensionless constant and the parameter $B$ is no
longer free. The coupling constant
$\theta(\tilde{X}_{\nu})$ is plotted in Fig.1.
For simplicity, in the following we write the constraint relation
into
\begin{align}
r_+=\theta\times B. \label{rb}
\end{align}

The scalar field $\phi(r)$ at the black hole
horizon becomes
\begin{eqnarray}
\phi(r_+)=\pm\sqrt{\frac{8B}{r_{+}+B}}=\pm\sqrt{\frac{8}{1+\theta}},\label{phi}
\end{eqnarray}
which is independent of the horizon radius $r_+$.

%%%%%%%%%%%%%%%%%%%%%%%%%%%%%%%%%%%%%%%%%%%%%%%%%%%%%%%%%%%%%%%%%%%%%%%%%%%
\begin{figure}[h]
\includegraphics{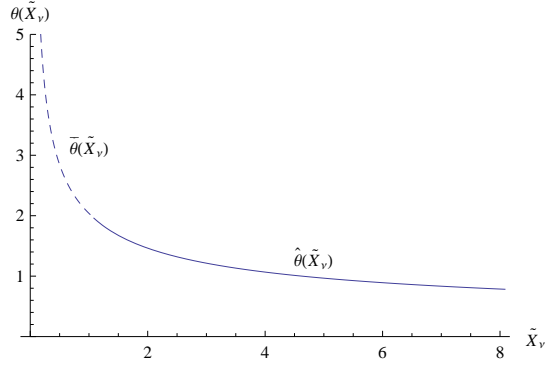}
\caption{$\theta(\tilde{X}_{\nu})$ vs. $\tilde{X}_{\nu}$.}
\end{figure}

\section{Thermodynamics of rotating black hole with nonminimal scalar hair}
\label{3s}

In this rotating hairy black hole, the mass and
angular momentum can be calculated by adopting
the Brown-York method \cite{Brown:1992br}. The
quasilocal mass $m(r)$ at  $r$ takes the form
\cite{Brown:1994gs,Creighton:1995au,Chan:1995wj}
\begin{eqnarray}
m(r)=\sqrt{f(r)}E(r)-j(r)\omega(r),\label{3b}
\end{eqnarray}
where $E(r)=2(\sqrt{f_0(r)}-\sqrt{f(r)})$ is the
quasilocal energy at $r$ and
$j(r)=\frac{d\omega(r)}{dr}r^3$ is the quasilocal
angular momentum. Here, $f_0(r)$ is a background metric function
that determines the zero of the energy. By setting $B=Q=0$, we can obtain
the background metric $f_0(r)=\frac{r^2}{l^2}$.
As a result, the mass and
angular momentum of the black hole can be
obtained as
\begin{eqnarray}
M\equiv \lim_{r\rightarrow\infty}m(r)=-3\mu B^2,\quad
J\equiv \lim_{r\rightarrow\infty}j(r)=6\alpha B^2.\label{4b}
\end{eqnarray}
Thus the black hole metric coefficient
[Eq.~(\ref{nf})] can be rewritten into
\begin{eqnarray}
f(r)=-M\left(1+\frac{2B}{3r}\right)+\frac{r^2}{l^2}+\frac{\left(3r+2B\right)^2J^2}{36r^4},\quad
\omega(r)=-\frac{\left(3r+2B\right)J}{6r^3}.\label{f}
\end{eqnarray}
The condition $-\mu\geq\frac{2\alpha}{l}$ is
needed to protect the cosmic censorship, which
puts the constraint $\frac{M}{J}\geq\frac{1}{l}$
for this black hole.

\subsection{First law of black hole thermodynamics}

With Eq.~(\ref{rb}), the mass of the rotating
hairy black hole reads
\begin{eqnarray}
M&=&\frac{J^2l^2\left(2B+3r_+\right)^2+36r_+^6}{12l^2r_+^3\left(2B+3r_+\right)}\nonumber\\
&=&\frac{J^2l^2\left(2+3\theta\right)^2+36r_+^4\theta^2}{12l^2r_+^2\theta\left(2+3\theta\right)},\label{nm}
\end{eqnarray}
and the entropy is
\begin{eqnarray}
S&=&\frac{A_H}{4G}\left(1-\xi \phi^2(r_+)\right)\nonumber\\
&=&\frac{4\pi r_+^2}{B+r_+}=\frac{4\pi\theta r_+}{1+\theta}.\label{s}
\end{eqnarray}
The temperature of this rotating hairy black hole
can be derived as
\begin{eqnarray}
T&=&\frac{f'(r_+)}{4\pi}=\frac{(B+r_+)\left[36r_+^6-J^2l^2(2B
+3r_+)^2\right]}{24\pi l^2 r_+^5\left(2B+3r_+\right)}\nonumber\\
&=&\frac{(1+\theta)\left[36r_+^4\theta^2-J^2l^2(2
+3\theta)^2\right]}{24\pi l^2 r_+^3\theta^2\left(2+3\theta\right)}.\label{nt}
\end{eqnarray}
For $T=0$, we can obtain the radius of extremal rotating hairy black hole
$r_{ext}=\left[\frac{Jl(2+3\theta)}{6\theta}\right]^{1/2}$.

In addition, there exist two commuting Killing
vector fields for the metric [Eq.~(\ref{met})]
\begin{eqnarray}
\xi_{(t)}=\frac{\partial}{\partial t}, \quad \xi_{\psi}=\frac{\partial}{\partial \psi}.\label{12b}
\end{eqnarray}
The various scalar products of these Killing
vectors can be expressed through the metric
components
\begin{eqnarray}
\xi_{(t)}\cdot\xi_{(t)}&=&g_{tt}=-f(r),\nonumber\\
\xi_{(t)}\cdot\xi_{(\psi)}&=&g_{t\psi}=r^2\omega(r),\nonumber\\
\xi_{(\psi)}\cdot\xi_{(\psi)}&=&g_{\psi\psi}=r^2.\label{13b}
\end{eqnarray}
To examine physical processes near such a black
hole, we introduce a family of locally
nonrotating observers. The  angular velocity for
these observers that move on orbits with constant
$r$ and with a 4-velocity $u^{\mu}$ satisfying
$u\cdot\xi_{(\psi)}=0$ is given by
\cite{Aliev:2007qi,Yue:2011et}
\begin{eqnarray}
\Omega=-\frac{g_{t\psi}}{g_{\psi\psi}}=-\omega(r)=\frac{\left(3r+2B\right)J}{6r^3}.\label{14b}
\end{eqnarray}
When approaching the black hole horizon, the
angular velocity $\Omega_H$ turns out to be
\begin{eqnarray}
\Omega_H&=&-\omega(r_+)=\frac{\left(3r_{+}+2B\right)J}{6r_+^3}\nonumber\\
&=&\frac{(2+3\theta)J}{6\theta r_+^2}.\label{oh}
\end{eqnarray}
Combining these quantities, $M$, $T$, and $\Omega_H$, we
can verify that the first law of thermodynamics
holds in this case,
\begin{eqnarray}
dM=TdS+\Omega_H dJ
\end{eqnarray}
and the Smarr relation can be found,
\begin{eqnarray}
M-\Omega_{H} J=\frac{1}{2}TS.
\end{eqnarray}

In general, the local thermodynamic stability is
determined by the specific heat. The positive
specific heat guarantees the stability of the
black hole. When the specific heat becomes
negative, it indicates that the black hole will
be destroyed when it encounters a small
perturbation. The specific heat can be calculated
through $C_J=T\left(\frac{\partial S}{\partial
T}\right)_J$, which reads
\begin{eqnarray}
C_J=\frac{32\pi^2 l^2r_+^4\theta^3(2+3\theta)T}{\left(1+\theta\right)^2\left[
\left(2+3\theta\right)^2J^2l^2+12r_+^4\theta^2\right]}.\label{16b}
\end{eqnarray}
Apparently,  $C$ is always  positive when $T>0$,
which implies that the rotating hairy black hole
is locally stable when $r_+>r_{ext}$.

\subsection{Phase transition between the rotating black hole with nonminimal
scalar hair and rotating BTZ black hole}

Now we examine the behavior of free energy $F$,
which can be calculated as
\begin{eqnarray}
F=M-TS=\frac{J^2l^2\left(2+3\theta\right)^2-12r_+^4\theta^2}{4l^2r_+^2\theta\left(2+3\theta\right)}.\label{FF}
\end{eqnarray}
When the black hole horizon
$r_+=r_c=\left(\frac{Jl\left(2+3\theta\right)}{2\sqrt{3}\theta}\right)^{1/2}$,
the free energy vanishes. At $r_c$, the black
hole temperature $T=T_c$, where
\begin{eqnarray}
T_c=\frac{J(1+\theta)}{3^{1/4}l\pi}\sqrt{\frac{\theta}{2(2+3\theta)J l}}.
\end{eqnarray}
The critical temperature $T_c$ depends on the
values of $\theta$. The $F-T$ relation
with different values of $\theta$ is plotted in  Fig.~2,
where we see that  $T_c$ increases with the growth
of $\theta$. Moreover, $F$ changes its sign at temperature $T_c$.
In the region of $0<T<T_c$, the free energy of this hairy black hole
is always positive so the black hole phase evaporates completely to
a pure thermal radiation phase. When $T>T_c$,
the free energy of this hairy black hole will be less than
that of pure thermal radiation, and then this hairy
black hole will be thermodynamically favored. Hence there exist the
Hawking-page phase transition between the black hole with nonminimally
coupled scalar hair and the pure thermal radiation \cite{Hawking:1982dh}.

%%%%%%%%%%%%%%%%%%%%%%%%%%%%%%%%%%%%%%%%%%%%%%%%%%%%%%%%%%%%%%%%%%%%%%%%%%%
\begin{figure}[h]
\includegraphics{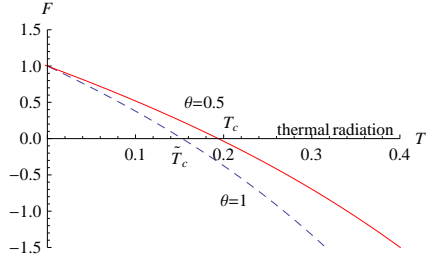}
\caption{The free energy $F$ of the rotating
hairy black hole relates to the temperature $T$
for different $\theta$ when we fix $J=1$ and
$l=1$.}
\end{figure}

It is worth comparing the free energies for the
hairy rotating black hole and the BTZ black hole.
It is necessary to point out that, for a fixed value of $M$,
from Eq.~(\ref{f}) we see that the scalar field parameter $B$ cannot be
switched off naively to reduce the hairy black
hole into the BTZ black hole. The rotating hairy black
hole and the BTZ black hole cannot be smoothly
deformed into each other; they belong to
different disconnected phases.

In the limit $\phi=0$, the scalar potential
$V(\phi)$ reduces to $-\frac{1}{l^2}$, and the
action, Eq.(1), admits the rotating BTZ black hole \cite{Banados:1992wn}
\begin{eqnarray}
ds^2=-\left(-\hat{M}+\frac{\rho^2}{l^2}+\frac{\hat{J}^2}{4\rho^2}\right)dt^2
+\left(-\hat{M}+\frac{\rho^2}{l^2}+\frac{\hat{J}^2}{4\rho^2}\right)^{-1}d\rho^2+
\rho^2\left(d\psi-\frac{\hat{J}^2}{2\rho^2}dt\right)^2.\label{17b}
\end{eqnarray}
The thermodynamic quantities, such as the temperature $\hat{T}$, mass $\hat{M}$,
entropy $\hat{S}$, and free energy $\hat{F}$ of the rotating
BTZ black hole are given by
\begin{eqnarray}
&&\hat{M}=\frac{\rho_+^2}{l^2}+\frac{\hat{J}^2}{4\rho_+^2}, \quad
\hat{T}=\frac{4\rho_+^4-\hat{J}^2l^2}{8\pi l^2\rho_+^3}, \quad
\hat{S}=4\pi\rho_+, \nonumber\\
&&\hat{F}=\frac{3\hat{J}^2l^2-4\rho_+^4}{4l^2\rho_+^2},\quad
\Omega_H=\frac{\hat{J}}{2\rho_+^2}.\label{18b}
\end{eqnarray}

To compare the free energies of the hairy black
hole and the BTZ black hole,  we need to match
the temperature $T=\hat{T}$ and the angular
momentum $J=\hat{J}$ of these two black holes,
\begin{eqnarray}
\frac{(1+\theta)\left[36r_+^4\theta^2-J^2l^2(2
+3\theta)^2\right]}{24\pi l^2 r_+^3\theta^2\left(2+3\theta\right)}
=\frac{4\rho_+^4-J^2l^2}{8\pi l^2\rho_+^3}.\label{19b}
\end{eqnarray}
In Fig.~3(a), we plot the free energies $F$
of rotating hairy and BTZ black holes for
different values of $\theta$.
We find that the free energies $F$ of
rotating hairy black holes are always larger than
that of the rotating BTZ black hole when $T>0$.
This means that thermodynamically the rotating BTZ black
hole phase is more thermodynamically preferred.
There exists a possible thermodynamical phase
transition for the hairy black hole to become
a rotating BTZ black hole, provided that there are some
thermal fluctuations. On the other hand,
we also evaluate the free energies of both
black holes for different values of $J$; see Fig.3(b).
The free energy $\hat{F}$ of a rotating BTZ black hole is
always larger than that of the rotating hairy black hole.
It implies that the relationship between the free energies of the
rotating BTZ and hairy black holes is not
affected by the values of the angular momentum
$J$.

%%%%%%%%%%%%%%%%%%%%%%%%%%%%%%%%%%%%%%%%%%%%%%%%%%%%%%%%%%%%%%%%%%%%%%%%%%%
\begin{figure}[htb]
  \subfigure[$J=1$]{\label{fig:subfig:a} %% label for second subfigure
  \includegraphics{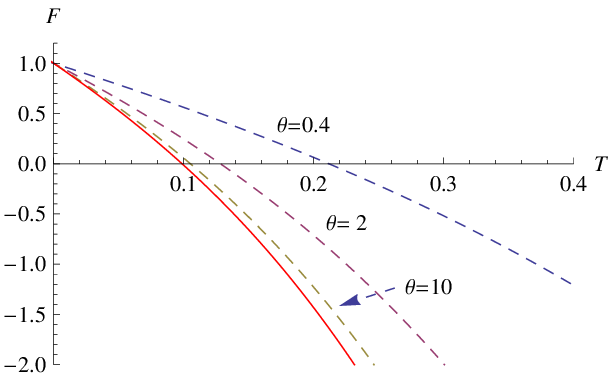}}%
  \hfill%
  \subfigure[$\theta=0.2$]{\label{fig:subfig:b} %% label for second subfigure
  \includegraphics{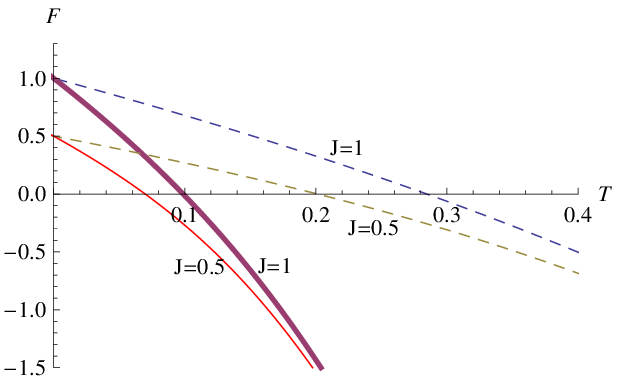}}%
  \caption{ The free energies $F$ of (2+1)-dimensional rotating hairy black hole (dashed lines)
and BTZ black hole (solid lines) vs the temperature $T$ with $l=1$. }
\end{figure}

\section{Thermodynamics of Rotating black hole with minimally coupled scalar hair}
\label{4s}

In this section, we generalize the above
discussions to the three-dimensional rotating
black hole with minimally coupled scalar hair.
This black hole was obtained
from the action \cite{Xu:2014uha}
\begin{align}
I=\frac{1}{2}\int \mathrm{d}^{3}x\sqrt{-g}\left(
R- \nabla_{\mu} \phi\nabla^{\mu} \phi -2V(\phi )\right),
\end{align}
where $\phi(r)$ is the scalar field and the scalar potential
takes the form
\begin{align}
    V(\phi)&=-\frac{1}{l^2}\cosh^6\left(\frac{1}{2\sqrt{2}}\phi\right)
    +\frac{1}{l^2}\left(1+\mu l^2\right)
    \sinh^6\left(\frac{1}{2\sqrt{2}}\phi\right)\nonumber\\
    &-\frac{\alpha^2}{64}\sinh^{10}\left(\frac{1}{2\sqrt{2}}\phi\right)
    \cosh^{6}\left(\frac{1}
    {2\sqrt{2}}\phi\right)\bigg[\tanh^6\left(\frac{1}{2\sqrt{2}}\phi\right)
    \nonumber\\
    &-5\tanh^4\left(\frac{1}{2\sqrt{2}}\phi\right)
    +10\tanh^2\left(\frac{1}{2\sqrt{2}}\phi\right)-9\bigg]
\end{align}
with the constant parameters $l$, $\mu$, and $\alpha$.
Then, the rotating hairy black hole solution is obtained as \cite{Xu:2014uha}
\begin{align}
\mathrm{d}s^{2}&=-\left( \frac{H(r)}{H(r)+B}\right) ^{2}f(H(r))\mathrm{d}t^{2}
+\left( \frac{H(r)+B}{H(r)+2B}\right) ^{2}\frac{\mathrm{d}r^{2}}{f(H(r))}\nonumber\\
&+r^{2}\bigg(\mathrm{d}\psi-\omega(H(r))\mathrm{d}t\bigg) ^{2},\\
&f(H(r))=3\mu B^2+\frac{2\mu B^3}{H(r)}+\frac{\alpha^2 B^4(3H(r)+2B)^2}{H(r)^4}
+\frac{H(r)^2}{l^2},\\
&\omega(H(r))=\frac{\alpha B^2(3H(r)+2B)}{H(r)^3},
\quad H(r)=\frac{1}{2}\left(r+\sqrt{r^{2}+4B r}\right)
\end{align}
and the scalar field is described by
\begin{align}
\phi(r) =2\sqrt{2}{\rm arctanh}\sqrt{\frac{B}{H(r)+B}}.
\end{align}

The event horizon is located at \cite{Xu:2014uha}
\begin{align}
H(r_+)=h\times B,
\end{align}
where the parameter $h$ only takes the real
positive value of the three roots of
$f(H(r_+))=0$
\begin{align}
  &h^{(1)}=\frac{X_1}{\tilde{X}_0^{(i)}}+\frac{1}{X_1},\nonumber\\
  &h^{(2,3)}=-\frac{1}{2}\bigg(\frac{X_1}{\tilde{X}_0^{(i)}}
  +\frac{1}{X_1}\bigg)
  \pm I\frac{\sqrt{3}}{2}\bigg(\frac{X_1}{\tilde{X}_0^{(i)}}
  -\frac{1}{X_1}\bigg) \label{H_+}
\end{align}
with
\begin{align}
  &X_1= \left[\left(1+\sqrt{\frac {-1+\tilde{X}_0^{(i)}}{\tilde{X}_0^{(i)}}}
  \right)\left(\tilde{X}_0^{(i)}\right)^{2}\right]^{1/3},\quad (i=1,2), \\
&X_0^{(1)}=-\,{\frac {\mu\ell+\sqrt {{\mu}^{2}{\ell}^{2}
  -4\,{\alpha}^{2}}}{2{\alpha}^{2}\ell B^2}}=\frac{\tilde{X}_0^{(1)}}{B^2},\\
&X_0^{(2)}=-\,{\frac {\mu\ell-\sqrt {{\mu}^{2}{\ell}^{2}
-4\,{\alpha}^{2}}}{2{\alpha}^{2}\ell B^2}}=\frac{\tilde{X}_0^{(2)}}{B^2}.
\end{align}
We require $\mu\leq-\frac{2\alpha}{\ell}$  to
satisfy $X_0^{{i}}>0, (i=1,2)$.

The thermodynamical quantities of the rotating
black hole with minimally coupled scalar field
are given by \cite{Xu:2014uha}
\begin{align}
&M=\,{\frac {{J}^{2}{l}^{2}(3h+2)^2+36h^2\,H(r_+)^{4}}{12{l}^{2}h \left( 3h+2\right)H(r_+)^{2} }},\quad
T=\frac{(1+h)\left[36H(r_+)^4h^2-J^2l^2(2
+3h)^2\right]}{24\pi l^2h^2\left(2+3h\right) H(r_+)^3},\nonumber\\
&S=\frac{4\pi h H(r_+)}{(h+1)}, \quad \Omega_H=\frac{(3h+2)J}{6h H(r_+)^3},\quad
F=\,{\frac {\,{J}^{2}{l}^{2}(3h+2)^2-12h^2\,H(r_+)^{4}}{4{l}^{2}h \left(3h+2\right)H(r_+)^{2}}}.
\end{align}
These expressions are similar to their
counterparts
[Eqs.~(\ref{nm}), (\ref{s}), (\ref{nt}), (\ref{oh}), and (\ref{FF})]
for the rotating black hole nonminimally coupled
with scalar field. It is easily found that the
first law of black hole thermodynamics is well
protected in this black hole background. Moreover,
the free energy $F$ of this rotating hairy black hole changes its sign
at the critical temperature
$T_c=\frac{J(1+h)}{3^{1/4}l\pi}\sqrt{\frac{h}{2(2+3h)J l}}$.
It is argued in Ref.~\cite{Hawking:1982dh} that
this rotating hairy black hole is thermodynamically favored in the region of $T>T_c$,
and when $0<T<T_c$, this rotating hairy black hole phase evaporates
completely into the pure thermal radiation.
Therefore, there exists the so-called Hawking-Page phase transition between the
black hole with the minimally coupled scalar field phase and the pure
thermal radiation phase.

Following the above discussions, we can further
examine the phase transitions between the
rotating BTZ and this kind of rotating hairy
black holes for the same temperature and angular
momentum values:
\begin{align}
\frac{(1+h)\left[36H(r_+)^4h^2-J^2l^2(2
+3h)^2\right]}{24\pi l^2h^2\left(2+3h\right) H(r_+)^3}
=\frac{4\rho_+^4-J^2l^2}{8\pi l^2\rho_+^3}.
\end{align}
We find that the similar phenomena occur
for these free energies; see Fig.~4.
Here, the rotating BTZ black hole always possesses smaller free energy
than the rotating black hole with a minimally coupled
scalar field, which means that the rotating BTZ black hole
is a more thermodynamically preferred phase. The thermodynamical
behavior we observed here is very similar to the
rotating black hole counterpart with nonminimally
coupled scalar field.

%%%%%%%%%%%%%%%%%%%%%%%%%%%%%%%%%%%%%%%%%%%%%%%%%%%%%%%%%%%%%%%%%%%%%%%%%%%
\begin{figure}[h]
\includegraphics{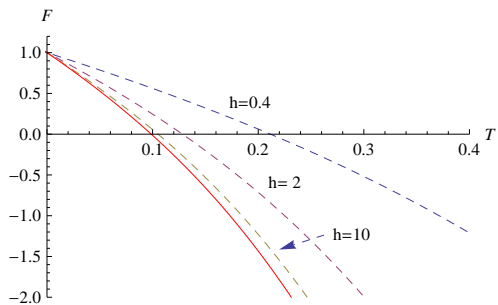}
\caption{The free energies $F$ of
(2+1)-dimensional rotating black hole with
minimally scalar field (dashed lines) and BTZ
black hole (solid line) vs the temperature
$T$ when $l=1$. }
\end{figure}

\section{Conclusions and Discussions}
\label{5s}

In the backgrounds of three-dimensional
rotating black holes with a nonminimally coupled scalar field,
we found that when the scalar field parameter $B$ is arbitrary,
the first law of thermodynamics cannot be protected.
By adopting a proper form of scalar potential $V(\phi)$,
a proper form of rotating black hole solution in
the three-dimensional Einstein gravity with a nonminimally coupled
scalar field have been obtained, in which
the scalar field parameter $B$ is constrained to
relate to the black hole size $r_+$, and the first law of
thermodynamics of black hole is satisfied in this case.

We have further calculated the free energy in the
three dimensional rotating hairy black hole
backgrounds. Comparing the free energies of this
rotating hairy black hole and the pure AdS space, we found
that the Hawking-Page phase transition exists
between this rotating hairy black hole and the pure AdS
space. Moreover, we have computed the free
energies for the rotating hairy black hole
and the rotating BTZ
black hole when they have the same temperature
and angular momentum. We disclosed that the rotating BTZ black hole has smaller
free energy which is a thermodynamically more
preferred phase.
This property is general no matter whether the
rotating black hole is minimally coupled or
nonminimally coupled to scalar field.

In addition, some new rotating hairy black hole solutions have
been recently found, such as the charged hairy black hole
\cite{Xu:2013nia,Cardenas:2014kaa}, the
Born-Infeld hairy black hole
\cite{Mazharimousavi:2014vza}, the rotating charged
hairy black hole with infinitesimal electric
charge and rotation parameters
\cite{Sadeghi:2013gmf}, and
black hole dressed by a (non)minimally coupled scalar
field in new massive gravity
\cite{Correa:2014ika}, etc. It would be
interesting to extend our discussion to explore
the thermodynamical properties and phase
transitions of these new black hole solutions and
see whether the results obtained here are
general.

{\bf Acknowledgments}

This work was supported by the National Natural Science Foundation of China.
D.C.Z is extremely grateful to Shao-Jun Zhang, Xiao-Mei Kuang and Cheng-Yong Zhang for useful
discussions.

\end{document}